\documentclass[11pt, twoside]{article}

\usepackage{newpasp,epsf}
\def\Msun{M_\odot}

\begin{document}

\title{A Massive Dark Object in the S0 galaxy NGC 4350}
\author{E. Pignatelli}
\affil{SISSA/ISAS, via Beirut 4, I-34014 Trieste, Italy}

\begin{abstract}
We present a new detection of a Massive Dark Object in
the S0 Galaxy NGC 4350, obtained applying a new
dynamical model on ground-based photometric and
kinematic data already present in literature. 
\end{abstract}

\noindent
There is increasing evidence that most galaxies with a
large spheroidal component host a Massive Dark Object
(MDO) in their centre, with masses ranging from $\sim
10^8 \Msun$ to $2 \cdot 10^{10} \Msun$ (Ho, 1999).

The goal of this work is to investigate the possible
presence of an MDO in the nucleus of the S0 galaxy NGC
4350, by applying a new dynamical model (Pignatelli \&
Galletta, 1999) to ground-based data already present in
literature (Seifert and Scorza, 1996; Fisher, 1997). The
galaxy was selected on the basis of the high observed
rotational velocity of the nuclear disk.

We show that the photometric profiles and the kinematics
along the major and minor axes, including the $h_3$ and
$h_4$ profiles, imply the presence of a central MDO of
mass \hbox{$M_{\rm MDO} \sim 1.5 - 9.7 \cdot 10^{8}
\Msun$}, i.e. $0.3-2.8$\% of the mass derived for the
stellar spheroidal component. Models without MDO are
unable to reproduce the kinematic properties of the inner
stars and of the rapidly rotating nuclear gas.

It is interesting to compare the value we found with the
known relations between the masses of the central black
holes and the observed ``central'' stel-
\vspace{-0.2cm}
\begin{figure}[h]
\plotfiddle{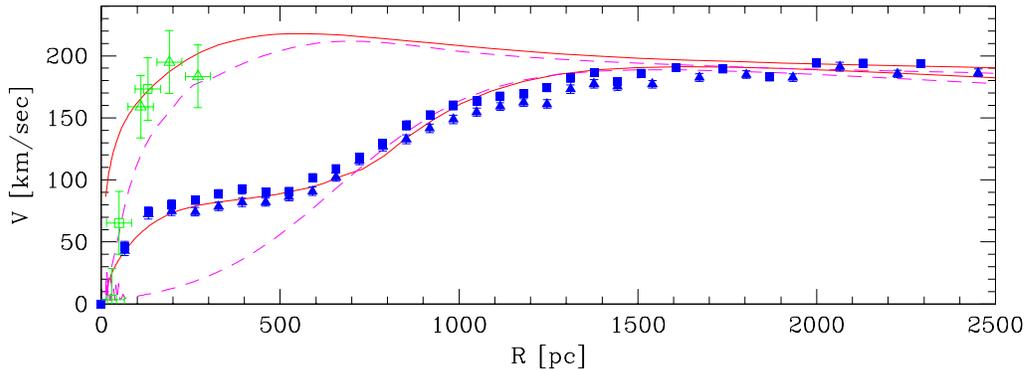}{4.6cm}{0}{70}{70}{-230}{-120}
\caption{Stellar ({\em filled symbols}) and gas ({\em open
 symbols}) rotational velocities (Fisher,1997) for the major axis,
 compared with our best-fit model circular and velocities
 with ({\em solid line}) and without ({\em long-dashed
 line}) a central MDO. The squares and triangles
 represent data respectively from the approaching and the 
 receding side. 
}
\end{figure}

\begin{figure}[ht]
\plotfiddle{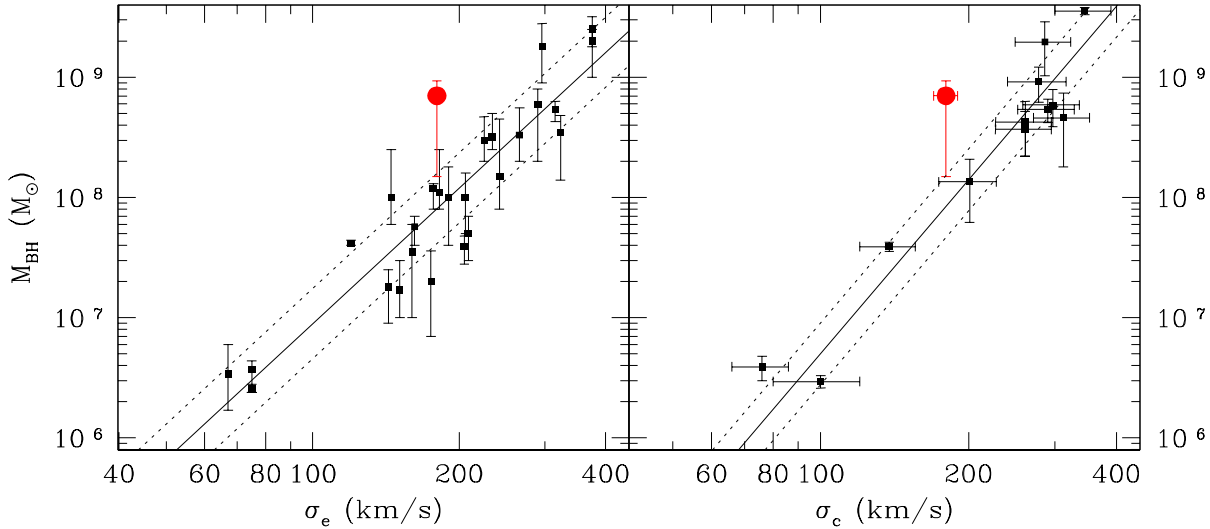}{6.8cm}{0}{80}{80}{-260}{-350}
\caption{Comparison between the values of $\sigma_e$ and
$M_{\rm MDO}$ we found for NGC 4350 (filled circle) and the relation
given by Gebhardt et al. (2000; left plot) and Ferrarese
\& Merritt (2000; right plot). }
\end{figure}

\noindent
lar velocity
dispersions (see Fig. 2). While the best-fit
given by the stellar kinematics is almost 10 times higher
than the average relation, the lower limit we find is
consistent (within the observed scatter) with this
relation. On the other hand, 
we expect (because of selection effects) this
galaxy to host a MDO a bit more massive than the average
for this velocity dispersion. Moreover, it is known that
ground-based stellar kinematics -if the systematic errors
are not properly taken into account- gives $M_{\rm MDO}$
higher than the ones obtained with HST spectroscopy. In
this work, when all the possible sources of errors are    
taken into account, we find results which are again
consistent with the HST average values (see Pignatelli et
al., 2000 for a complete discussion).
              
We also found that the stellar population of this galaxy
comprise of an exponential disc (27\% of the light) and a
diffuse spheroidal component (73\% of the light) that
cannot be represented by a simple de Vaucouleurs profile
at any radius. The $M/L$ ratios we found for the stellar
components (respectively 3.3 and 6.6) are typical of
those of disc and elliptical galaxies.

\end{document}